# Single-shot implementation of dispersion-scan for the characterization of ultrashort laser pulses

D. Fabris[1], W. Holgado[2], F. Silva[3,4], T. Witting[1], J. W. G. Tisch[1,5], and H. Crespo[3]

[1]Blackett Laboratory, Imperial College, London SW7 2AZ, UK
[2]Grupo de Investigación en Óptica Extrema (GIOE), Universidad de Salamanca, E-37008 Salamanca, Spain
[3]IFIMUP-IN e Departamento de Física e Astronomia, Faculdade de Ciências, Universidade do Porto, 4169-007 Porto, Portugal
[4]*fsilvaportugal@gmail.com*
[5]*john.tisch@imperial.ac.uk*

**Abstract:** We demonstrate a novel, single-shot ultrafast diagnostic, based on the dispersion-scan (d-scan) technique. In this implementation, rather than scanning wedges to vary the dispersion as in standard d-scan, the pulse to be measured experiences a spatially varying amount of dispersion in a Littrow prism. The resulting beam is then imaged into a second-harmonic generation crystal and an imaging spectrometer is used to measure the two-dimensional trace, which is analyzed using the d-scan retrieval algorithm. We compare the single-shot implementation with the standard d-scan for the measurement of sub-3.5-fs pulses from a hollow core fiber pulse compressor. We show that the retrieval algorithm used to extract amplitude and phase of the pulse provides comparable results, proving the validity of the new single-shot implementation down to near single-cycle durations.

## 1. Introduction

In the ultrafast domain, optical pulse characterisation poses a significant challenge [1]. Basic autocorrelation techniques only give limited and incomplete information about the pulse duration [2], especially with the advent of shorter and more complex pulses that possess structured spectra and non-trivial spectral phase profiles. Moreover, such complex pulses are exploited in experiments and applications heavily dependent on the precise details of the pulse waveform and carrier-envelope phase, such as attoscience [3] and high-field physics [4]. Hence, there is growing demand for accurate and reliable pulse diagnostics capable of dealing with complicated pulses with durations down to the single-cycle or even sub-cycle regime.

Different approaches have been developed to characterize broadband pulses by retrieving both amplitude and/or phase information. The two most common measurement techniques are frequency resolved optical gating (FROG) [5] and spectral phase interferometry for direct electric field reconstruction (SPIDER) [6], with all their respective variants – see for example [7–10]. These are extremely versatile and powerful techniques, but they involve non-trivial operations, such as temporal overlap of short pulses, beam splitting and recombination and achieving adequate phase-matching bandwidth in the required nonlinear processes. Additional challenges arise when measuring few-cycle and octave-spanning pulses due to the need to accommodate their ultra-broad bandwidths in all previous steps [11]. The dispersion-scan (d-scan) technique [12, 13], a new technique for the simultaneous measurement and compression of femtosecond laser pulses, has been originally proposed as a way to simplify such steps, making both the setup and the data acquisition process as simple as possible while still providing robust and precise pulse retrieval. In fact it involves only the measurement of a spectrum multiple times, without major manipulations required on the laser pulse. The d-scan technique however is not as studied as more established techniques like FROG and SPIDER, which warrants further development. For this reason several experiments have already been performed with d-scan on various pulses [12–15]. The combination of the d-scan and STARFISH techniques has enabled the spatiotemporal and spatio-spectral characterization of sub-two-cycle laser pulses, both collimated and on focus [16]. Third-harmonic variants of d-scan have also been successfully demonstrated based on multilayer graphene [17] and on thin films with variable bandgaps [18]. Comparative studies between d-scan and FROG [19] and d-scan and SPIDER [20] have also been performed using few-cycle pulses, revealing excellent agreement between techniques.

So far, single-shot d-scan is yet to be demonstrated and this is the main focus of the present work. Single-shot capability is of great interest, not only for low repetition rate lasers, where the nature of the experiments is inherently based on the shot-to-shot characterization of the system, but also for higher repetition rate systems where the reproducibility of the pulses may be achieved via feedback loops or where data-tagging techniques are employed.

## 2. Concept and experiment

In the standard d-scan [12], the second-harmonic of the pulse under investigation is generated in a nonlinear crystal and the resulting spectrum is measured as a function of the introduced dispersion using a 1-D spectrometer. The complete measurement is built up over a large number of shots as the dispersion is scanned, usually by varying the amount of glass insertion using a motorized wedge pair. Pulse characterization through d-scan is based on the fact that when a pulse undergoes a nonlinear conversion process, such as second-harmonic generation (SHG), the resulting spectral intensity has a well-defined dependence on the input spectral phase. By measuring the spectrum of the nonlinear signal for different input phases (glass insertions) around the point of maximum compression, a two-dimensional trace is obtained that enables the full retrieval of the spectral phase of the pulses via an iterative algorithm. Additionally, d-scan has a totally inline and robust setup, without the need of beam splitting, beam recombination or interferometric precision.

For a single shot implementation of d-scan we aim to acquire the two-dimensional trace required by the d-scan algorithm without the need for moving parts. The concept of our single-shot implementation is depicted in Fig. 1. A spatially variable amount of dispersion is imparted on the beam profile using a glass prism. The resulting SHG signal is re-imaged into a thin SHG crystal using reflective optics. The resulting SHG signal is then re-imaged onto a broadband imaging spectrometer with minimized aberrations [21], which provides a spatially resolved second-harmonic spectrum, $S(x, \lambda)$ [7], where each spectral slice at a given x corresponds to a second-harmonic spectrum obtained with a different amount of dispersion.

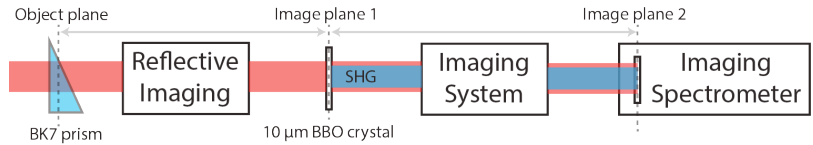

Fig. 1. Schematic of the single-shot d-scan concept and implementation.

The experimental setup is shown in Fig. 2. The laser system used in the experiment is the CPA system (Femtolasers GmbH) and Ar-filled hollow core fiber (HCF) pulse compressor described in [14], which includes a water cell for third-order dispersion compensation. The output of this system, which is capable of producing pulses in the single-cycle regime with energies up to 200 µJ, could be directed both to a standard (scanning) d-scan setup and to the single-shot implementation with a flip mirror, in order to achieve a direct comparison of the two diagnostics. The standard d-scan setup is based on that in [14], but employs a 10 µm thick BBO crystal (instead of 5 µm) for SHG, and a broadband wire grid polariser (Thorlabs, Inc.) to separate the second-harmonic signal from the fundamental (instead of the previous scheme comprising a spatial mask, Brewster-angled reflection and slit). The thicker BBO crystal is important for achieving a high signal-to-noise ratio in the single-shot system, which uses the same nonlinear crystal as the standard d-scan. Although the 10 µm thick crystal has a narrower phase matching bandwidth compared to the 5 µm crystal, this does not prevent the d-scan algorithm from fully retrieving the pulses. This stems from the intrinsic robustness of d-scan with respect to bandwidth limitations in the measured second-harmonic signal, an important property presented in [12] and further demonstrated with few-cycle pulses [13, 14].

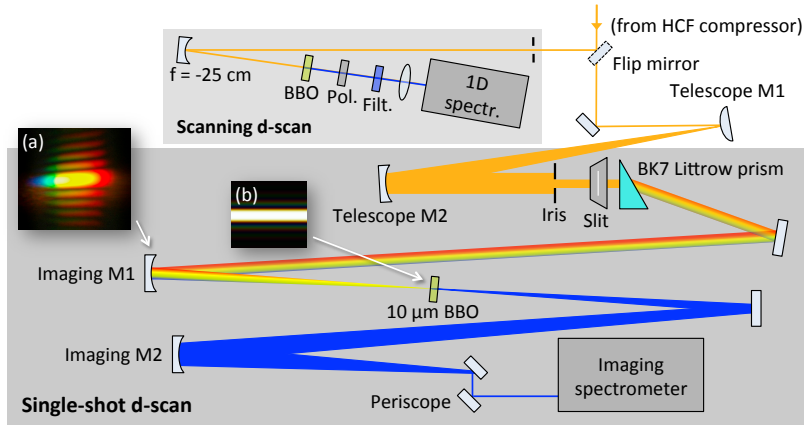

Fig. 2. Optical setup, comprising the standard scanning d-scan setup (top) and the new single-shot d-scan implementation (bottom) – see text for more details.

The beam directed to the single-shot d-scan was expanded with a telescope (magnification factor M = -3), with two spherical mirrors positioned such that astigmatism was avoided [22]. This expansion is necessary to obtain a spatially uniform beam profile, by selecting the central portion of the expanded beam in order to minimise spatio-temporal couplings [16] that are not under investigation in this experiment. An iris was used to select a beam diameter of approximately 11 mm. The beam was then transmitted through a 400 μm horizontal slit and directed to a N-BK7 Littrow prism with 30-60-90º angles (Edmund Optics Ltd.) to introduce a spatially varying dispersion along the beam. The size of the beam and the apex angle of the prism allowed for a glass insertion difference between the two extrema of the beam of 5 mm (corresponding to a group-delay dispersion range of approximately 220 $fs^2$), which is sufficient to have a confined and retrievable d-scan trace for the near-single-cycle pulse under investigation [14]. The output plane of the Littrow prism was re-imaged by a focusing mirror ($f$ = 45 cm) onto a BBO crystal (10 μm thick, 1 cm square, cut for type-I SHG at 795 nm), with a magnification factor M = -0.33. Light transmitted though the slit and prism combination becomes spatially chirped at the plane of the focusing mirror due to angular dispersion in the prism. This is shown in the inset (a) of Fig. 2, which also displays wavelength-dependent diffraction from the slit. Nevertheless, the image produced by the focusing mirror at the plane of the thin BBO crystal is free from such effects (inset (b) of Fig. 2). Ray tracing analysis showed that this system effectively provides aberration free imaging. The focusing also ensures sufficient intensity at the BBO crystal (estimated at $65 \pm 10$ GW/cm$^2$ for maximum pulse compression) to obtain a detectable second-harmonic signal and allow us to collect the full spatial information, with a linear correspondence between spatial position on the prism surface and spatial position on the BBO. This was achieved by re-imaging the SHG from the BBO crystal onto the entrance slit of the imaging spectrometer by a second focusing mirror ($f$ = 50 cm). In this way the spatially dependent dispersion was mapped to the spatial axis of the imaging spectrometer.

The calibration of the dispersion axis in the case of the scanning d-scan is straightforward. Once the wedge pair is positioned perpendicularly to the incoming beam, the stage step size can be directly converted to the corresponding glass insertion amount by simple trigonometry. In the case of the single shot d-scan the dispersion axis calibration was performed in a similar way. A pair of motorized wedges was positioned in the beam and different amounts of glass were inserted. The effect on the trace measured with the imaging spectrometer was a linear position shift in the spatial axis direction, as expected [12], thus further confirming the expected linear correspondence of different amounts of dispersion to different positions. From this information the pixel-to-insertion calibration can be performed.

## 3. Results and analysis

Two datasets are shown. The first one is the standard d-scan, performed by recording different SHG spectra as a function of wedge insertion. The second one is the trace measured in single-shot mode with the imaging spectrometer. The two datasets were taken consecutively by removing the flip mirror shown in Fig. 2. Due to the reduced sensitivity of the camera in the ultraviolet (UV) region, we opted for integration times of 200 ms in order to increase the signal-to-noise ratio for our proof of concept measurement. Nevertheless, the diagnostic is intrinsically scan free and potentially capable of measuring a single pulse.

The measured and retrieved traces are shown in Fig. 3. The single-shot measurement could not cover the same spectral region as the scanning one, especially for the shorter wavelengths, due to the poor spectral response of the CCD camera in the UV.

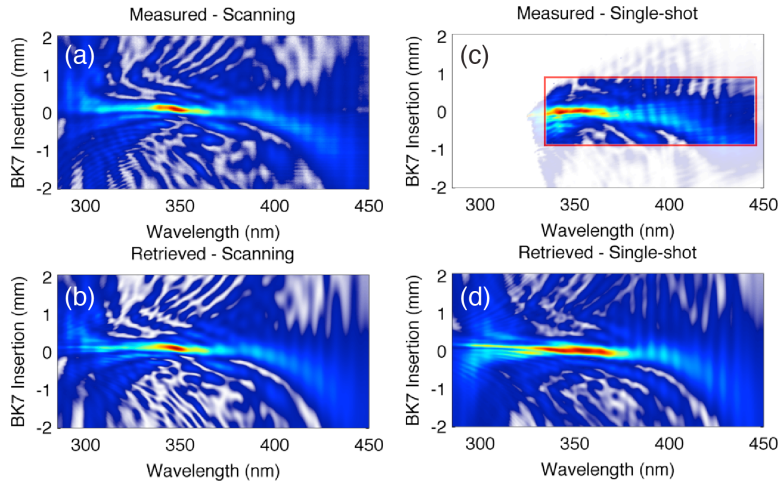

Fig. 3. Experimental d-scan traces measured with the standard scanning mode (a), and with the single shot setup (c). The corresponding algorithm results are shown in (b) and (d), respectively.

We further limited the portion of the single-shot traces used in the retrieval to the area within the red rectangle in Fig. 3(c) so as to exclude regions with poor signal-to-noise ratio. Again, this is not too detrimental for retrieval, which can cope with partial traces, as shown in [14] and demonstrated here by the similarity between the retrieved traces for both setups in Figs. 3(b) and (d) – note the trace features outside the main pulse that are faithfully retrieved by the algorithm from the single-shot trace, even though they are not part of the measurement.

Since the windows over which the signal was considered were different for each method, it is not meaningful to directly compare the corresponding retrieval errors (as defined in [12]). Nonetheless, the error over the full represented window is 2% for the scanning measurement and 9% for the single-shot case – the difference mainly due to the missing signal in the latter.

The pulses retrieved from the traces are shown in Fig. 4. The input spectrum necessary for the retrieval algorithm is shown in Fig. 4(a) (Fourier limit of 2.8 fs), while in Fig. 4(b) the two retrieved spectral phases are presented, showing an overall good agreement over the main portion of the spectrum, from 530 to 990 nm. The blue curves correspond to 16 independent retrievals performed from the scanning trace under different initial conditions, while the black curve is the spectral phase retrieved from the single-shot measurement (average of 16 independent retrievals); the grey area is the corresponding standard deviation. In both methods, we see that the higher-order phase is less well retrieved than the average phase, as evidenced by the opposing rapid phase oscillations over the central portion of the spectrum, from 700 to 800 nm, and the higher rms errors associated with them. The temporal results are shown in Fig. 4(c). The relatively small insertion and wavelength ranges of the single-shot trace compared to the scanning trace increase the difference in the pre- and post-pulse structure around the main pulse, but the results from both methods are nevertheless comparable. The agreement between the two main parts of the temporal profiles is very good, with the pulse duration measured with the scanning and the single shot setups resulting in 3.2 ± 0.1 fs and 3.3 ± 0.1 fs, respectively.

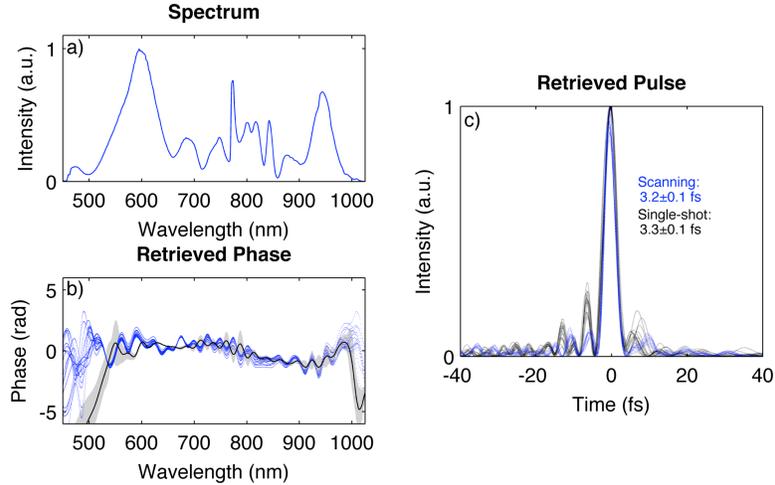

Fig. 4. Pulses retrieved from the two d-scan setups: (a) measured spectrum; (b) comparison of the retrieved spectral phase obtained with the scanning technique (blue curves, 16 independent retrievals) and the single-shot setup (black curve). The grey area is the error on the single-shot retrieval (based on 16 independent retrievals). The corresponding pulses are shown in (c) with the same color code.

## 4. Conclusions

We have presented a single-shot implementation of the d-scan pulse characterisation technique. A spatially dependent amount of dispersion was applied to the beam with a Littrow prism, and then the spatially resolved second-harmonic spectrum was recorded with an imaging spectrometer. Such a setup is able to measure, in a single shot, the d-scan trace of the pulse. We measured a near-single-cycle pulse from an Ar-filled hollow core fiber pulse compressor and compared the results of the standard wedge-scanning implementation with data from the single-shot implementation. The agreement between the spectral phase retrieved in the two configurations is very good, resulting in 3.2 ± 0.1 fs in the scanning measurement and 3.3 ± 0.1 fs in the single-shot measurement, proving the validity of the new single-shot implementation down to near single-cycle pulse durations. The single-shot implementation makes the benefits of the d-scan technique available to laboratories with lower repetition rate laser systems or to those using feedback or data-tagging techniques where diagnostics that average over multiple shots are unsuitable.

**Acknowledgments**

We gratefully acknowledge financial support from the European Science Foundation through the SILMI programme (Short Visit Grant 5585), the UK Engineering and Physical Sciences

Research Council (grants EP/F034601/1 and EP/I032517/1) and the Portuguese funding agency, FCT - Fundação para a Ciência e Tecnologia (grant PTDC/FIS/122511/2010), co-funded by COMPETE and FEDER. H. Crespo and F. Silva gratefully acknowledge support from FCT grants SFRH/BSAB/105974/2015 and SFRH/BD/69913/2010, respectively.